\documentclass[12pt]{article}   %%LaTeX2e
\usepackage{epsfig}
\newcommand{\gsim}{\;\rlap{\lower 3.5 pt \hbox{$\mathchar \sim$}} \raise 1pt
 \hbox {$>$}\;}
\newcommand{\lsim}{\;\rlap{\lower 3.5 pt \hbox{$\mathchar \sim$}} \raise 1pt
 \hbox {$<$}\;}
\newcommand{\equ}[1]{Eq.\,(\ref{#1})}

\newcommand{\non}{\nonumber}
\newcommand{\beq}{\begin{equation}}
\newcommand{\eeq}{\end{equation}}
\newcommand{\bea}{\begin{eqnarray}}
\newcommand{\eea}{\end{eqnarray}}
\newcommand{\smallw}{{\scriptscriptstyle W}} %

\newcommand{\mw}{M_\smallw}
\newcommand{\muw}{\mu_\smallw}
\newcommand{\msq}{\tilde{m}}

\newcommand{\mst}[1]{m_{\tilde{t}_{#1}}}

\newcommand{\tb}{\tan \beta}

\def \ct  {c_{\tilde{t}}}
\def \st  {s_{\tilde{t}}}

\def\as{\alpha_s}

\def \mt   {m_t}

\def \mg   {m_{\tilde{g}}}

\def\slash#1{\setbox0=\hbox{$#1$}#1\hskip-\wd0\dimen0=5pt\advance
       \dimen0 by-\ht0\advance\dimen0 by\dp0\lower0.5\dimen0\hbox
         to\wd0{\hss\sl/\/\hss}}
\def\gequiv{\raise 0.4ex \hbox{$>$} \kern -0.7 em \lower 0.62 ex \hbox{$\sim$}}
\def\gappeq{\mathrel{\rlap {\raise.5ex\hbox{$>$}}
{\lower.5ex\hbox{$\sim$}}}}

\def\pl#1#2#3{{ Phys. Lett. }{\bf B#1~}(19#2)~#3}
\def\zp#1#2#3{{ Z. Phys. }{\bf C#1~}(19#2)~#3}
\def\prl#1#2#3{{ Phys. Rev. Lett. }{\bf #1~}(19#2)~#3}

\def\pr#1#2#3{{ Phys. Rev. }{\bf D#1~}(19#2)~#3}
\def\np#1#2#3{{ Nucl. Phys. }{\bf B#1~}(19#2)~#3}

\begin{document}              

%%%%%%%%%%%%%%%%%%%%%%%%%%%%%%% titlepage %%%%%%%%%%%%%%%%%%%%%%%%%%%%%%%%%%%%

\begin{titlepage}
\begin{flushright}
        \small
        CERN-TH/2000-283\\
        DFPD-00/TH/43\\
        September 2000  
%\\        hep-ph/0009337
\end{flushright}

\begin{center}
\vspace{1cm}

{\large\bf $B\to X_s \gamma$ in supersymmetry: large contributions\\
beyond the leading order}

\vspace{0.5cm}
\renewcommand{\thefootnote}{\fnsymbol{footnote}}
{\bf    G.~Degrassi$^a$,
        P.~Gambino$^b$,
        G.F.~Giudice$^b$\footnote
                {On leave of absence from INFN, Sez. di Padova, Italy.}}
\setcounter{footnote}{0}
\vspace{.8cm}

{\it
        $^a$ Dipartimento di Fisica, Universit{\`a}
                  di Padova, Sezione INFN di Padova,\\ 
                  Via F.~Marzolo 8 , I-35131 Padua, Italy\\
\vspace{2mm}
        $^b$ Theory Division, CERN CH-1211 Geneva 23, Switzerland \\
\vspace{1cm} }

{\large\bf Abstract}
\end{center}

\vspace{.5cm} 
We discuss possible large contributions to $B\to X_s \gamma$, which can occur
at the next-to-leading order in supersymmetric models. They can originate from
terms enhanced by $\tan\beta$ factors, when the ratio between the two Higgs
vacuum expectation values is large, or by logarithm of $M_{susy}/M_W$, when the
supersymmetric particles are considerably heavier than the $W$ boson. We give
compact formulae which  include 
 all potentially large higher-order contributions.
We find that  $\tan\beta$ terms at the next-to-leading order do not only appear
from the Hall-Rattazzi-Sarid effect (the modified relation between the bottom
mass and Yukawa coupling), but also from an analogous effect in the top-quark
Yukawa coupling. Finally, we show how  next-to-leading order corrections, in
the large $\tan\beta$ region, can significantly reduce the limit on the
charged-Higgs mass, even if supersymmetric particles are very heavy.

\noindent

% PACS numbers:

\end{titlepage}
\newpage
\eject

\section{Introduction}

The inclusive radiative decay $B \to X_s \gamma$ provides a powerful
experimental testing ground for physics beyond the Standard Model,
because of its sensitivity to new particle virtual effects. The 
measurements of $B \to X_s \gamma$ at CLEO, LEP, and Belle~\cite{cleo},
and the progress in the precision expected from experiments at the
$B$ factories requires a substantial effort in limiting the uncertainty
in the theoretical calculation. 
This programme has been carried out
in the Standard Model up to next-to-leading order
corrections~\cite{nlo1,nlo2,noi,nlo3}, 
reducing the theoretical error in the prediction
for the branching ratio of $B \to X_s \gamma$ to about 10\%, equally
distributed between renormalization scale dependence 
and uncertainties 
in the input parameters~\cite{paolo}. 
Moreover, non--perturbative effects appear to be under control 
\cite{nonpert} and several refinements have been introduced in the
analysis \cite{paolo,refine}.
In the case of the two-Higgs doublet models, complete next-to-leading order 
analyses have been presented in ref.~\cite{noi,BG}. Although the leading-order
contributions to $B \to X_s \gamma$ in  supersymmetric models are well
studied~\cite{tutti}, the situation at the next-to-leading order has not
been fully analyzed. The next-to-leading order QCD corrections have been
calculated~\cite{noi2,misiak} assuming minimal flavour violation~\cite{gabr} 
({\it i.e.} 
assuming that the Cabibbo-Kobayashi-Maskawa matrix is the only source
of flavour violation at the weak scale) and choosing a hierarchical
spectrum in which charginos and one stop are lighter than gluinos and
the other squarks. In a complementary approach, a systematic 
leading-order analysis of the contributions from flavour-violating gluino
exchanges has been presented in ref.~\cite{BGHW}, in a very interesting and 
complete study.

In this paper we want to follow a different approach. We will not try to
compute the complete set of next-to-leading corrections in a general
supersymmetric model (a rather formidable task), but instead we will
identify all potentially large two-loop corrections in models with
minimal flavour violation.
These come in two
classes: {\it (i)} corrections enhanced by $\tan\beta$, in the case
in which the ratio of the two Higgs vacuum expectation values is
large~\cite{rat2};
 {\it (ii)} corrections enhanced by a logarithm of the ratio $\mu_{SUSY}
/\mu_W$, in the case in which the scale of supersymmetric particles
$\mu_{SUSY}$ is much larger than the scale $\mu_W$ of the $W$ or top
mass.
In this way, we obtain analytic formulae which, we believe, will give
a very good approximation of a complete calculation and, because of 
their simple form, are very practical to be implemented in analyses
of supersymmetric models.

The paper is organized as follows. In sect.~2 we discuss the terms enhanced
by $\tan\beta$ in the next-to-leading corrections to the Standard Model
and charged Higgs contributions to the relevant Wilson coefficients. In
the limit of very heavy supersymmetric particles, there are two sources of
such terms in the charged Higgs contribution. One is coming from the 
finite corrections to the bottom quark mass (the so-called 
Hall-Rattazzi-Sarid effect~\cite{hall}), while the other one
is related to its counterpart for the top quark \cite{noi2}.
 In sect.~3 we describe
the $\tan^2\beta$ terms appearing at two--loops in the chargino contribution
and the log--enhanced contributions in the terms subleading in $\tan\beta$.
Section~4 contains a summary of the formulae 
for the large higher--order contributions %two-loop terms of 
to the Wilson coefficients; these formulae can be directly
implemented in  analyses for $B \to X_s \gamma$. Some numerical results
of such an analysis are illustrated in sect.~5. In particular, we 
show how  next-to-leading order corrections, in
the large $\tan\beta$ region, can significantly reduce the limit on the
charged-Higgs mass, even if supersymmetric particles are very heavy. 

\section{Standard Model and charged Higgs contributions}

In this paper we are focusing on short-distance contributions and, therefore,
we can restrict our discussion to the form of the Wilson coefficients
of the $\Delta B=1$ magnetic and chromo-magnetic operators $Q_7=(e/16\pi^2)m_b
{\bar s}_L \sigma^{\mu \nu}b_RF_{\mu\nu}$ and
$Q_8=(g_s/16\pi^2)m_b
{\bar s}_L \sigma^{\mu \nu}t^ab_RG^a_{\mu\nu}$ evaluated at the matching
scale $\mu_W$ in the effective Hamiltonian:
\beq
{\cal H}= -\frac{4G_F}{\sqrt{2}}V_{ts}^*V_{tb} \sum_i C_i(\mu_W) Q_i(\mu_W).
\eeq
 At the leading order, the contributions to
$C_7(\mu_W)$ and $C_8(\mu_W)$ from the Standard Model particles and
from the charged Higgs boson are given by
\bea
C_{7,8}^{(SM)}(\mu_W)&=& F_{7,8}^{(1)}(x_t) \\
C_{7,8}^{({H^\pm})}(\mu_W)&=& \frac1{3\, \tan^2 \beta} F_{7,8}^{(1)}(y_t) +
  F_{7,8}^{(2)}(y_t)\label{chiggs}
\eea
where
\beq
x_t=\frac{\bar m_t^2(\muw )}{\mw^2},~~~~~~~~~~~~~
y_t=\frac{\bar m_t^2(\muw )}{M_H^2}~.
\eeq
Here $\bar m_t^2(\muw )$ is the SM running top mass 
and
\bea
F_7^{(1)}(x)&=&\frac{x(7-5x-8x^2)}{24(x-1)^3}+\frac{x^2(3x-2)}{4(x-1)^4}\ln x
\label{f71}\\
F_8^{(1)}(x)&=&\frac{x(2+5x-x^2)}{8(x-1)^3}-\frac{3x^2}{4(x-1)^4}\ln x
\label{f81}\\
F_7^{(2)}(y)&=&\frac{y(3-5y)}{12(y-1)^2}+\frac{y(3y-2)}{6(y-1)^3}\ln y \\
F_8^{(2)}(y)&=&\frac{y(3-y)}{4(y-1)^2}-\frac{y}{2(y-1)^3}\ln y~.
\eea
\
The relation between the Wilson coefficients at $\mu_W$ and the
branching ratio for $B\to X_s \gamma$ is well known
(see for example refs.\cite{nlo1,noi}).

The charged Higgs contribution of eq.(\ref{chiggs}) 
 consists of two terms. In the large
$\tan \beta$ limit, the first one (in which the chiral flip occurs on
the external bottom quark line) is suppressed by $1/\tan^2\beta$,
while the second one (in which the chiral flip occurs in the
charged Higgs vertex) is independent of $\tan \beta$. The absence
of a term enhanced by $\tan\beta$ is a consequence of the fact that,
in the large $\tan\beta$ limit, $H^\pm$ decouples from the
right-handed top quark. This property is not maintained at the
next order in perturbation theory in a supersymmetric model, 
and thus we expect two-loop charged-Higgs
contributions to $C_7$ and $C_8$ enhanced by $\tan\beta$.

Let us now extract the $\tan \beta$-enhanced terms. At one-loop, the
relation between the bottom quark mass $m_b$ and Yukawa coupling $y_b$
receives a finite correction proportional to $\tan \beta$~\cite{hall}:
%(see also \cite{carena})
\beq
m_b=\sqrt{2}M_W~\frac{y_b}{g}~\cos\beta \left( 1+\epsilon_b \tan\beta
\right) .
\label{massab}
\eeq
The coefficient $\epsilon_b$, generated by gluino-sbottom and 
chargino-stop diagrams, is given by~\cite{hall}
\beq
\epsilon_b=-\frac{2\,\as}{3\,\pi} \frac{\mu}{\mg} 
H_2(x_{\tilde{b}_1\,\tilde{g}},x_{\tilde{b}_2\,\tilde{g}}) -
           \frac{ y_t^2}{16\, \pi^2} \,\tilde{U}_{a 2}\frac{A_t}{m_{\chi^+_a}}
\,H_2(x_{\tilde{t}_1\,\chi^+_a},x_{\tilde{t}_2\,\chi^+_a}) 
\, \tilde{V}_{a 2}~.
\eeq
For simplicity, we have not explicitly written down the other 
weak contributions
to $\epsilon_b$, which can be found in ref.~\cite{pierce}.
Here $A_t$ is the trilinear coefficient, 
$\tilde{U}$ and $\tilde{V}$ are the two matrices (assumed to be real)
that diagonalize the chargino mass matrix according to 
%($M_2$ is the weak gaugino mass)
\beq
\tilde{U} \pmatrix{ M_2 & \mw \sqrt{2} \sin \beta \cr
              \mw \sqrt{2} \cos \beta & \mu } \tilde{V}^{-1}\, 
\eeq
and
\beq
H_2(x,y)  =   
\frac{x\ln\, x}{(1-x)(x-y)} + \frac{y\ln\, y}{(1-y) (y-x)}~.
\eeq
Here and in the following we define, for generic indices $\alpha$
and $\beta$,
\beq 
x_{\alpha \, \beta}  \equiv \frac{m^2_\alpha}{m^2_\beta}.
\eeq
The analogous contribution to the top quark mass ($\epsilon_t$)
is irrelevant for us, since it gives rise to terms
suppressed by $\tan\beta$. 

The effective
Yukawa couplings of the charged Higgs current-eigenstates $H_D^+$
and $H_U^+$ (belonging to the doublets coupled to down- and up-type
quarks, respectively) are given by
\beq
{\cal L}= \sum_{d}  V_{t d} \,y_t
{\bar t}_R d_L \left[ H_U^+ +\epsilon_t^\prime (d)\, H_D^+\right] 
- \sum_{u} V_{u b} \,
y_b {\bar u}_L b_R \left[ H_D^+ +\epsilon_b^\prime (u) \, H_U^+\right] 
+{\rm h.c.}
\label{lagr}
\eeq
The sum $\sum_u \,(\sum_d)$ is over the 3 generation of
up (down) type quarks and 
we have kept only the terms proportional to third-generation Yukawa
couplings.
\begin{figure}[t]%\vspace{-2.7cm}
\begin{center}
\mbox{\epsfxsize=11cm\epsffile{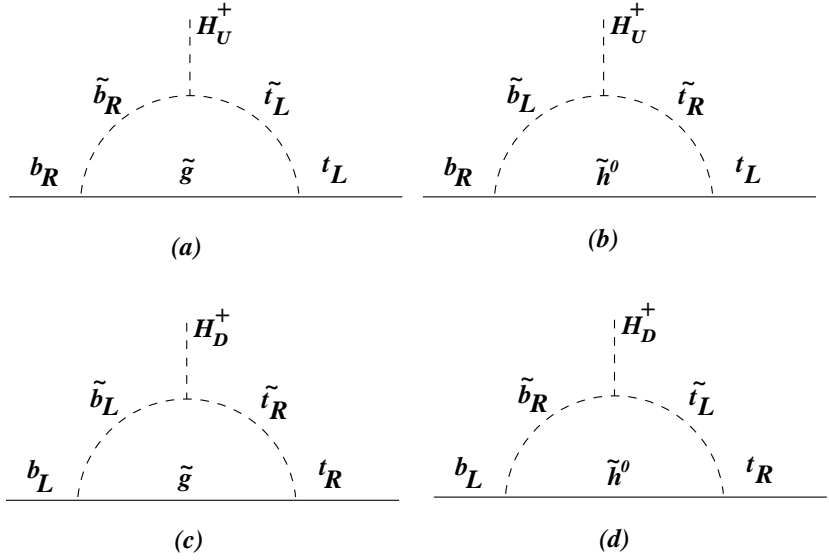}}
\end{center}
\vspace{-.3cm}
\caption{\sf Feynman diagrams (for current squark eigenstates)
representing  the QCD (a) and Yukawa
(b)  contributions to $\epsilon_b^\prime (t)$ and the QCD (c) and
Yukawa (d) contribution to $\epsilon_t^\prime (b)$. 
\label{fig0}}
\end{figure}
The $\epsilon_{b,t}^\prime$ coefficients originate from the charged-current 
analogue of the diagrams leading to $\epsilon_{b,t}$ --- see Fig.1 ---
 and they are given
by
\bea
\epsilon_b^\prime (t)&=& -\frac{2\,\as}{3\,\pi} \frac{\mu}{\mg}
   \left[ \ct^2 c^2_{\tilde{b}}
\,H_2(x_{\tilde{t}_1\,\tilde{g}},x_{\tilde{b}_2\,\tilde{g}}) + 
\ct^2 s^2_{\tilde{b}}
\,H_2(x_{\tilde{t}_1\,\tilde{g}},x_{\tilde{b}_1\,\tilde{g}}) + \right. 
\nonumber \\ 
&& \left. ~~~~~~~~~~~~~~~ \st^2 c^2_{\tilde{b}} \,  
H_2(x_{\tilde{t}_2\,\tilde{g}},x_{\tilde{b}_2\,\tilde{g}}) 
+ \st^2 s^2_{\tilde{b}} \,  
H_2(x_{\tilde{t}_2\,\tilde{g}},x_{\tilde{b}_1\,\tilde{g}})
\right]  \nonumber \\
&& - \frac{ y_t^2}{16\, \pi^2} \,N_{4 a}\frac{A_t}{m_{\chi^0_a}}
   \, \left[ \ct^2 c^2_{\tilde{b}}\,
    H_2(x_{\tilde{t}_2\,\chi^0_a},x_{\tilde{b}_1\,\chi^0_a}) + 
    \ct^2 s^2_{\tilde{b}}\,
    H_2(x_{\tilde{t}_2\,\chi^0_a},x_{\tilde{b}_2\,\chi^0_a}) + \right. 
\nonumber \\
&&\left.~~~~~~~~~~~~~~~\st^2 c^2_{\tilde{b}}\,
    H_2(x_{\tilde{t}_1\,\chi^0_a},x_{\tilde{b}_1\,\chi^0_a}) + 
    \st^2 s^2_{\tilde{b}}\,
    H_2(x_{\tilde{t}_1\,\chi^0_a},x_{\tilde{b}_2\,\chi^0_a}) \right]
             \, N^*_{a 3} \\
%%%%%%%%%%%%%%%%%%%%%%%%%%%%%%%%%%%%%%%
\epsilon_t^\prime (b)&=& -\frac{2\,\as}{3\,\pi} \frac{\mu}{\mg}
\left[ \ct^2 c^2_{\tilde{b}}
\,H_2(x_{\tilde{t}_2\,\tilde{g}},x_{\tilde{b}_1\,\tilde{g}}) + 
\ct^2 s^2_{\tilde{b}}
\,H_2(x_{\tilde{t}_2\,\tilde{g}},x_{\tilde{b}_2\,\tilde{g}}) + \right. 
\nonumber \\ 
&& \left. ~~~~~~~~~~~~~~~ \st^2 c^2_{\tilde{b}} \,  
H_2(x_{\tilde{t}_1\,\tilde{g}},x_{\tilde{b}_1\,\tilde{g}}) 
+ \st^2 s^2_{\tilde{b}} \,  
H_2(x_{\tilde{t}_1\,\tilde{g}},x_{\tilde{b}_2\,\tilde{g}})
\right]  \nonumber \\
&& - \frac{ y_b^2}{16\, \pi^2} \,N^*_{4 a}\frac{A_b}{m_{\chi^0_a}}
   \, \left[ \ct^2 c^2_{\tilde{b}}\,
    H_2(x_{\tilde{t}_1\,\chi^0_a},x_{\tilde{b}_2\,\chi^0_a}) + 
    \ct^2 s^2_{\tilde{b}}\,
    H_2(x_{\tilde{t}_1\,\chi^0_a},x_{\tilde{b}_1\,\chi^0_a}) + \right. 
\nonumber \\
&&\left.~~~~~~~~~~~~~~~\st^2 c^2_{\tilde{b}}\,
    H_2(x_{\tilde{t}_2\,\chi^0_a},x_{\tilde{b}_2\,\chi^0_a}) + 
    \st^2 s^2_{\tilde{b}}\,
    H_2(x_{\tilde{t}_2\,\chi^0_a},x_{\tilde{b}_1\,\chi^0_a}) \right]
             \, N_{a 3} \label{pirl}
\eea
where $N$ is the matrix that diagonalizes
the neutralino mass matrix,  
$c_{\tilde{q}} \equiv \cos \theta_{\tilde{q}}$, $s_{\tilde{q}} 
\equiv \sin\theta_{\tilde{q}}$, and 
the squarks eigenstates are ${\tilde q}_1=c_{\tilde{q}} \,{\tilde q}_L
+ s_{\tilde{q}} \,{\tilde q}_R$ and 
${\tilde q}_2=- s_{\tilde{q}} \, {\tilde q}_L
+c_{\tilde{q}} \,{\tilde q}_R$ 
have mass eigenvalues $m_{{\tilde q}_1}>
m_{{\tilde q}_2}$.
The quantity 
$\epsilon_t^\prime (d)$ with $d\ne b$ is given by eq.~(\ref{pirl})
after setting $y_b=0$, $\theta_{\tilde b}=0$, and identifying
$m_{{\tilde b}_{1,2}}$ with $m_{{\tilde d}_{1,2}}$.
In the limit of exact weak
$SU(2)$ ({\it i.e.} ${\tilde m}_{b_L}= {\tilde m}_{t_L}$ and
$\theta_{\tilde t}=\theta_{\tilde b}=0$), the coefficients 
$\epsilon^\prime_{b,t}$ coincide
with $\epsilon_{b,t}$. However, their difference can be numerically
significant, since we are interested also in cases of large stop mixing.
Notice that the coefficients  $\epsilon^\prime_{b,t}$ (as well as
$\epsilon_{b,t}$) are non-vanishing even in the limit in which all
the supersymmetric masses are simultaneously sent to infinity. This means
that the effective charged-Higgs theory \cite{carena} obtained by 
decoupling gluino and 
squarks does not correspond to what is usually called Model II. There are
additional couplings, namely the $\epsilon$ and $\epsilon^\prime$ 
coefficients in eqs.~(\ref{massab}) and (\ref{lagr}), which lead to
$\tan\beta$-enhanced contributions to $C_7$ and $C_8$ at two loops. 
In the decoupling limit, $\epsilon^\prime_{b,t}-\epsilon_{b,t}$ vanish,
since they  are proportional to $SU(2)$-breaking effects. 

We can now express the interaction Lagrangian in eq.~(\ref{lagr}) 
in terms of the would-be Goldstone boson $G^+=\cos\beta  H_D^+ + \sin\beta
H_U^+$ and the physical charged Higgs $H^+=-\sin\beta H_D^++\cos\beta
H_U^+$. Replacing the Yukawa couplings $y_t$ and $y_b$ with the
top and bottom quark masses, see eq.~(\ref{massab}), we find
\bea
{\cal L}&=&\frac{g}{\sqrt{2}M_W} G^+ \left\{  
\sum_{d}m_t \, V_{t d} {\bar t}_R d_L
- \sum_{u} m_b \,V_{u b}
\frac{1+\epsilon^\prime_b (u)\tan\beta}{1 +\epsilon_b \tb}
{\bar u}_L b_R \right\} 
\nonumber \\
&+&\frac{g}{\sqrt{2}M_W} H^+ \left\{ 
\sum_{d}  V_{t d} 
\frac{m_t \left[ 1-\epsilon^\prime_t (d) \tan\beta\right] }{\tan\beta}
 {\bar t}_R d_L
+ \sum_{u} V_{u b} \frac{m_b \tan\beta}{1 +\epsilon_b \tan\beta} 
{\bar u}_L b_R \right\} \nonumber \\
&+&{\rm h.c.} 
\label{lagrr}
\eea
From the interactions in eq.~(\ref{lagrr}) we can now directly read
the leading $\tan\beta$  higher--order
 contributions to the Wilson coefficients $C_7$ and $C_8$ in
the case of a light charged Higgs, by
recalling that the one-loop diagram is proportional to the product
of the Higgs couplings to the ${\bar t}_Lb_R$ and ${\bar s}_Lt_R$
currents,
\bea
C_{7,8}^{(SM)}({\rm leading} \tan\beta)&=& 
\frac{\left[ \epsilon_b -\epsilon^\prime_b(t)\right]
\tan\beta}{1 +\epsilon_b\tan\beta}
 \, F_{7,8}^{(2)}(x_t) 
\label{tbSM} \\
C_{7,8}^{(H^\pm)}({\rm leading} \tan\beta)&=& 
-\frac{\left[ \epsilon^\prime_t(s) +\epsilon_b\right]
\tan\beta}{1 +\epsilon_b\tan\beta}
 \, F_{7,8}^{(2)}(y_t).
\label{tb2H}
\eea

In the limit of very heavy supersymmetric particles, the $\tan\beta$-enhanced
terms in $C_{7,8}^{(SM)}$ vanish, consistently with the theorem of
decoupling. However, for finite supersymmetric masses, we find
$\tan\beta$-enhanced corrections to the Standard Model contribution.

The $\tan\beta$-enhanced terms in the charged-Higgs contribution do
not vanish in the decoupling limit because, as previously explained,
additional couplings to those of the charged-Higgs Model II 
are recovered in this limit. The term proportional to $\epsilon_b$
(previously discussed in ref.~\cite{rat2})
originates from the modified relation between the bottom mass and
Yukawa coupling in eq.~(\ref{massab}), while the term proportional to
$\epsilon^\prime_t$ comes from the modified charged Higgs boson
vertex in eq.~(\ref{lagrr}) \cite{noi2}.

% A previous analysis~\cite{carena} of the effective charged Higgs
% Lagrangian has included the 
%corrections proportional to $\epsilon_b$, but missed the terms proportional
%to $\epsilon^\prime_t$.
%To circumvent the absence of $\tan\beta$ enhanced terms in the leading-order
%charged-Higgs contribution, one can consider,

Notice that  $\tan\beta$ enhanced terms in the charged-Higgs
contribution to $B\to X_s \gamma$
can be induced not only by  vertex corrections, 
but also by corrections to the charged-Higgs propagator.
%. Indeed,
%other $\tan\beta$ enhanced contributions to $B\to X_s \gamma$ can appear
They arise 
from loop corrections to the propagator $\langle H_U^+ H_D^-\rangle$
which vanishes at tree level, in the limit $\tan\beta \to \infty$.
In this limit, the Higgs potential classically has a Peccei-Quinn (PQ)
symmetry (since the scalar Higgs mixing mass parameter $B_\mu$ has
to vanish~\cite{hall}) and therefore any contribution to $\langle H_U^+ 
H_D^-\rangle$ must be proportional to $\mu$, which is the only surviving
PQ-breaking parameter. This propagator receives contributions  
from higgsino-gaugino
loops or stop loops. However, the loop contributions which renormalize
$B_\mu$ are simply reabsorbed in the definition of $\tan\beta$, which is
treated here as an input parameter. Analogously, the effect of PQ-violating
quartic Higgs couplings (of the kind $H_DH_UH_U^\dagger H_U$) obtained
by integrating out heavy supersymmetric particles (and which survive as
renormalizable interactions in the decoupling limit) are absorbed in the
redefinition of the vacuum expectation values. Nevertheless, supersymmetric
particles with masses comparable to $M_H$ can lead to momentum-dependent
contributions to $\langle H_U^+ H_D^-\rangle$, which generate  potentially
significant
$\tan\beta$ enhanced effects in $C_{7,8}$. In particular, this could be the 
case for the higgsino-gaugino loop, if charginos are light, although the
effect is suppressed by two powers of the $SU(2)_W$ breaking scale and
four powers of the weak gauge coupling constant. For this reason, we
will neglect this effect here. 

\section{Chargino contributions}
At the one-loop level, the leading
contribution to $C_{7,8}(\mu_W)$ in the large $\tan\beta$ limit from chargino
and squark exchange is given by
\bea
C_{7,8}^{\chi}(\muw)& =&
 \frac1{\cos\beta}\sum_{a=1,2}\left\{ 
\frac{\tilde{U}_{a2}\tilde{V}_{a1}\mw}{\sqrt{2} m_{\chi^+_a}}
\left[F^{(3)}_{7,8}(x_{\tilde{q}\, \chi^+_a})- \ct^2 \, 
F^{(3)}_{7,8} (x_{{\tilde{t}_1}\,\chi^+_a})
-\st^2 \,F^{(3)}_{7,8} (x_{{\tilde{t}_2}\,\chi^+_a}) \right]  \right.\non\\
&& 
+\left.\st \, \ct
\frac{\tilde{U}_{a2}\,
\tilde{V}_{a2} \,\bar{\mt}}{2\sin\beta \,m_{\chi^+_a}}
\left[ F^{(3)}_{7,8} (x_{\tilde{t}_1\,\chi^+_a}) -
F^{(3)}_{7,8} (x_{\tilde{t}_2\,\chi^+_a})\right]
 \right\} ~~~~,
\label{leading1}
\eea
where 
$m_{\tilde{q}}$ is a common mass for the first two generation squarks, and
\bea
F_7^{(3)}(x)&=&\frac{5-7x}{6(x-1)^2}+\frac{x(3x-2)}{3(x-1)^3}\ln x ,
\label{f73}\\
F_8^{(3)}(x)&=&\frac{1+x}{2(x-1)^2}-\frac{x}{(x-1)^3}\ln x~.
\label{f83}
\eea
\equ{leading1} contains two terms which grow linearly with $\tan\beta$, 
in the large $\tan\beta$ limit. 
%Their diagrammatic interpretation, in the mass
%insertion approximation, is shown in 
%fig.~1. 
The first line originates from
a chargino-stop loop, in which the
$\langle H_U\rangle$ insertion (signalling the $\tan\beta$ enhancement)
occurs in the gaugino-higgsino mixing. 
The second one, instead,  comes from a diagram mediated by a higgsino-stop 
loop and the
$\langle H_U\rangle$ insertion 
occurs in the stop left-right mixing. 
The first term suffers from a suppression with respect to the second one
of a $g^2/y^2_t$ factor (because of the gaugino coupling) and of
a squark GIM factor. However, in many cases of interest in which the stop
mass is significantly split from the $\tilde c_L$ and $\tilde u_L$ masses,
the two diagram contributions are comparable in size.

%The leading-order charged Higgs contribution discussed in the previous section
%has a constructive interference with the Standard Model contribution for
%any choice of $m_H$ and $\tan\beta$. The sign of the chargino contribution
%in eq.() is determined by ... {\it blabla sul segno.}

The only terms in the chargino contributions  enhanced by an extra
$\tan\beta$ factor 
at the two-loop order originate from the $\epsilon_b$ coefficient,
{\it i.e.} from the modified relation between $m_b$ and $y_b$ in 
eq.~(\ref{massab}). Therefore the leading higher--order terms are 
taken into account by dividing the one-loop expression of (20) by 
$(1+\epsilon_b \tan\beta)$ (see the first paper of \cite{rat2}).
Since the one-loop contribution is linear in
$\tan\beta$, the two-loop term is ${\cal O}(\tan^2\beta)$.

This method of extracting the leading $\tan\beta$ terms cannot give us the
two-loop ${\cal O}(\tan\beta)$ chargino contribution, which requires a
full diagrammatic calculation, outside the scope of this work. 
In the present analysis we are going to  keep
only the subleading terms in $\tan\beta$ which can be potentially large.
They are related to three different effects. {\it (i)} Two-loop 
${\cal O}(\tan\beta)$ terms which are power-suppressed only by the 
charged-Higgs mass and not by any supersymmetric particle heavy mass. 
These can be important in
models in which the charged-Higgs is significantly lighter than gluino 
and squarks, and have been computed in the previous section. {\it (ii)} 
Two-loop terms with large logarithms of the ratio $\mu_{SUSY}/\mu_W$
related to the different renormalizations of Yukawa couplings in the
Higgs/higgsino vertices. {\it (iii)} Two-loop terms enhanced by
$\ln(\mu_{SUSY}/\mu_W)$ connected with the anomalous dimensions of the
magnetic and chromo-magnetic effective operators. The last two classes
of terms become important when the scale of the
supersymmetric colored particles is
significantly higher than the $W$ and top masses. 
The effect is particularly sizeable because  the anomalous  dimensions
of the relevant operators are quite large. 

Let us now discuss the terms in class {\it (ii)}. The Yukawa coupling 
$\tilde{y}_t$ appearing in the chargino-stop-sbottom vertex 
is related to the ordinary top-quark Yukawa coupling $y_t$
only by supersymmetry. Therefore at the scale of the heavy supersymmetric
particle masses $\mu_{SUSY}$, we have 
$\tilde{y}_t (\mu_{SUSY})=y_t(\mu_{SUSY})$.
After decoupling the supersymmetric modes, $\tilde{y}_t$ is frozen, while
$y_t$ evolves according to the Standard Model renormalization group. Large
logarithms are generated when we express $\tilde{y}_t$ in terms of
$y_t$ or, in other words, in terms of the top quark mass evaluated at
the weak scale. 
The resummation of these logarithms gives
\bea
\tilde{y}_t (\mu_{SUSY}) &=& y_t (\mu_W) 
\left[ \frac{\alpha_s(\mu_{SUSY})}{\alpha_s(\mt)} \right]^{4/7}
\left[ \frac{\alpha_s(\mt)}{\alpha_s(\mu_W)} \right]^{12/23}\nonumber \\
& \times & \frac{1}{\sqrt{1+\frac{9y_t^2(m_t)}{8\pi\alpha_s(m_t)}
\left\{ \left[ \frac{\alpha_s(\mu_{SUSY})}{\alpha_s(\mt)} \right]^{1/7}-1
\right\} }}
\label{yina}
\eea
Here we have used six active quark flavors above the 
the top quark threshold which  is generally higher than $\mu_W$.
In practice, however,  $\mt=O(\mu_W)$ and the modification in the
running between $\mt$ and $\mu_W$ is numerically small. At worst,
for $\mu_W=40$ GeV it leads to a 1\% modification in the Yukawa coupling.
Therefore, the logarithms in class {\it (ii)} are taken into account by
using $\mt(\mu_{SUSY})$ in the chargino contribution.

The logarithms in class {\it (iii)} are taken into account by considering the
evolution of the effective operators from the scale $\mu_{SUSY}$ to the scale
$\mu_W$ (notice that no  new operator is involved)
\bea
C_7^\chi(\muw)&=& 
\eta^{-\frac{16}{3 \beta_0}} C_7^\chi(\mu_{SUSY})
+\frac83 \left(
  \eta^{-\frac{14}{3 \beta_0}}-\eta^{-\frac{16}{3 \beta_0}}\right) 
C_8^\chi(\mu_{SUSY})
\non\\
C_8^\chi(\muw)&=& 
\eta^{-\frac{14}{3 \beta_0}} C_8^\chi(\mu_{SUSY})
\label{c7rr}
\eea
where $\eta\equiv \alpha_s(\mu_{SUSY})/\alpha_s(\muw)=
[1 - (\beta_0/2 \pi) \ln(\mu_{SUSY}/ \mu_W)]^{-1}$ and 
$\beta_0 = -7$ corresponding to six active flavors. 

We can judge the effect of the resummation by retaining  only the first 
logarithm in eqs.~(\ref{c7rr})
\bea
\delta C_7^\chi(\muw)
&=&- \frac{4\alpha_s(\mu_W)}{3\pi} 
\left[ C_7^\chi(\mu_{SUSY}) -\frac13 C_8^\chi(\mu_{SUSY})
\right] \ln \frac{\mu^2_{SUSY}}{\muw^2}
\label{c7nr}\\
\delta C_8^\chi(\muw)
&=&- \frac{7\alpha_s(\mu_W)}{6\pi} C_8^\chi(\mu_{SUSY})
\ln \frac{\mu^2_{SUSY}}{\muw^2}.
\eea
For instance, taking $C_8^\chi(\mu_{SUSY})=0$ and $\mu_{SUSY}=1$ TeV, the 
not-resummed
evolutions of $C_7^\chi$ from $\mu_{SUSY}$ to $\mu_W$ is proportional to
\beq
- \frac{4 \as(\mu_W)}{3\pi}\ln
\frac{\mu_{SUSY}^2}{\muw^2}= -0.257; \ \ \ \  
- \frac{4 \as(\mu_{SUSY})}{3\pi}\ln
\frac{\mu_{SUSY}^2}{\muw^2}= -0.191; 
\eeq
while the resummed expression gives
\beq
\eta^{16/21}-1=-0.203~.
\eeq
This demonstrates that the choice of evaluating $\as$ at $\mu_{SUSY}$
can incorporate
most of the effect of the resummation.

There is another important case in which we can retain at next-to-leading 
order all terms 
linear in $\tan\beta$ not suppressed by heavy masses.
This is  when the charginos and a mostly right-handed
stop are significantly lighter than the gluino and the other squarks.
This scenario was discussed in ref.~\cite{noi2}, 
where  the leading terms
in an expansion in the ratio of light to heavy supersymmetric particle masses
were derived.
Also in that situation large logs (non--decoupling logs) 
dominate the next-to-leading corrections and may lead to sizeable effects in 
the branching ratio. It is not difficult to resum these logarithms.
The crucial point is that, in the
effective theory where only the gluino and heavy squarks have been 
integrated out at $\mu_{SUSY}$,
the gaugino and higgsino couplings renormalize differently from 
the ordinary gauge and Yukawa couplings.
As a result, the couplings in the chargino interactions 
at the electroweak scale differ from the Standard Model couplings
by $O(\alpha_s)$
contributions enhanced by large logarithms. If we denote by
$\tilde y$ and $\tilde g$ the Yukawa and weak gauge couplings in
the chargino vertex, supersymmetry requires ${\tilde y}(\mu_{SUSY})
=y (\mu_{SUSY})$ and ${\tilde g}(\mu_{SUSY})
=g (\mu_{SUSY})$. Since we evaluate the diagram involving charginos and light
stop at the scale $\mu_W$, we have to express the supersymmetric couplings
at the scale $\mu_W$ in terms of Standard Model couplings. Using 
the QCD renormalization group equations, we obtain
\beq
{\tilde g}(\mu_W)=g(\mu_W)\eta^{\frac{2}{\beta_0}},~~
{\tilde y}_t(\mu_W)=y_t(\mu_{SUSY})\eta^{\frac{2}{\beta_0}},~~
{\tilde y}_b(\mu_W)=y_b(\mu_W)\eta^{-\frac{2}{\beta_0}}
\eeq
where $\eta$ should be evaluated with $\beta_0=-41/6$,
the beta-function coefficient including 6 quark flavours  
and one squark. For convenience we have related ${\tilde y}_t(\mu_W)$
to the  top Yukawa coupling at $\mu_{SUSY}$, but ${\tilde y}_b(\mu_W)$
is expressed in terms of $y_b(\mu_W)$ in order to reconstruct the operators
$Q_7$ and $Q_8$ at the proper scale. The resummation of the QCD
logarithms is therefore implemented
 by evaluating $m_t$ at the scale $\mu_{SUSY}$ and
multiplying by $\eta^{4/\beta_0}$ the functions $F^{(1)}_{7,8}$ (which
describe diagrams proportional to two powers of ${\tilde g}(\mu_W)$ or
${\tilde y}_t(\mu_W)$). On the other hand, the functions $F^{(3)}_{7,8}$ 
should not be rescaled, because they arise from diagrams proportional
to ${\tilde y}_b(\mu_W){\tilde y}_t(\mu_W)$ or
${\tilde y}_b(\mu_W){\tilde g}(\mu_W)$. We will present the final
result in the next section. 

The large logarithms proportional to 
Yukawa couplings can also be resummed with the same approach we have
discussed here. However, in this case it is crucial to maintain the
electroweak gauge invariance of the effective theory between the scales
$\mu_{SUSY}$ and $\mu_W$, and therefore the resummation method is valid
only in the limit of pure right-handed light stop.

\section{Summary of the leading higher-order contributions}
\label{summary}
We now summarize the formulae for the supersymmetric contribution
to the Wilson coefficients that contain the leading higher-order effects
in the scenario in which the colored supersymmetric particles have mass
${\cal O}(\mu_{SUSY} \sim 1\, {\rm TeV})$ with the possibility 
that the physical charged Higgs and the charginos could be lighter with masses 
${\cal O}(\mu_W)$. The expressions,
up to next-to-leading order,  for the SM and charged Higgs contribution to 
$C_{7,8}$ can be found in ref.\cite{noi}. 
The next-to-leading order charged higgs 
contribution contains terms enhanced by potentially large logarithms
of the kind $\ln (m_H/\mu_W)$ which we have not resummed (see
 ref.~\cite{anlauf} for a complete resummation).\\
The chargino contribution is instead given
by eqs.(\ref{c7rr}) with
\bea
C_{7,8}^\chi(\mu_{SUSY})&=& 
 \sum_{a=1,2}\left\{ \frac2{3} \frac{\mw^2}{\msq^2} 
\tilde{V}_{a1}^2  F^{(1)}_{7,8}(x_{\tilde{q}\, \chi^+_a}) \right. \nonumber \\
&-& \frac2{3}  \left( \ct\,\tilde{V}_{a1}  - \st\, \tilde{V}_{a2} 
\frac{\bar{\mt}(\mu_{SUSY})}{\sqrt{2}\sin\beta\,\mw} \right)^2
\frac{\mw^2}{\mst{1}^2} F_{7,8}^{(1)}(x_{{\tilde{t}_1}\,\chi^+_a}) \nonumber \\
&-& \frac2{3}  \left( \st\,\tilde{V}_{a1}  + \ct\, \tilde{V}_{a2} 
\frac{\bar{\mt}(\mu_{SUSY})}{\sqrt{2}\sin\beta\,\mw} \right)^2 
\frac{\mw^2}{\mst{2}^2} F_{7,8}^{(1)}(x_{{\tilde{t}_2}\,\chi^+_a})\nonumber \\
&+&  \frac{K}{\cos\beta} \left(
\frac{\tilde{U}_{a2}\tilde{V}_{a1}\mw}{\sqrt{2} m_{\chi^+_a}}
\left[F^{(3)}_{7,8}(x_{\tilde{q}\, \chi^+_a})- 
\ct^2 \, F^{(3)}_{7,8} (x_{{\tilde{t}_1}\,\chi^+_a})
-\st^2 \,F^{(3)}_{7,8} (x_{{\tilde{t}_2}\,\chi^+_a}) \right] \right. \non\\
&+&\left.  \left. \st \, \ct 
\frac{\tilde{U}_{a2}\,
\tilde{V}_{a2} \,\bar{\mt}(\mu_{SUSY})}{2\sin\beta \,m_{\chi^+_a}}
\left[ F^{(3)}_{7,8} (x_{{\tilde{t}_1}\,\chi^+_a}) -
F^{(3)}_{7,8} (x_{{\tilde{t}_2}\,\chi^+_a})\right]
 \right) \right\} ~~.
\label{final}
\eea
In eq.(\ref{final}) $\bar{\mt}(\mu_{SUSY})$ is expressed in terms of the
top quark Yukawa coupling given by eq.(\ref{yina}) 
and the $K$ factor can be taken equal to 1 for 
small values of $\tb$, while in the large $\tb$ scenario is given by
$ K = 1/(1 + \epsilon_b \tb$). In the latter case one has to keep
the  contribution given by eq.(\ref{tbSM}) and, if the physical 
charged Higgs is assumed to be light, one can also take into account the 
dominant 
contribution to the terms enhanced by a single power of $\tb$ that is
given by eq.(\ref{tb2H}). 

We now consider  the second scenario in which charginos and the 
mostly right-handed
stop are significantly lighter than the gluino and the other squarks.
In this situation the NLO Wilson coefficients contain 
large logarithms of the ratio of light to heavy mass that
were not resummed in ref.\cite{noi2}, but can be easily included to
all orders using  the results of the previous section.
% and using the results in the appendix of the same paper, it is possible to
%resum them.

We identify  $\mu_{SUSY}$ with  the gluino mass scale, $\mg$,
 and $\mu_W$ with the charginos--light stop scale and
rewrite the supersymmetric contribution
to the Wilson coefficient at the weak scale  as 
\bea 
C_{7}^{\chi} (\mu_W) &=& 
 \sum_{a=1,2} \eta^{-\frac{16}{3 \beta_0}}
\left\{ \frac2{3} \frac{\mw^2}{\msq^2} 
\tilde{V}_{a1}^2  F^{(1)}_{7}(x_{\tilde{q}\, \chi^+_a}) \right. \nonumber \\
&-& \frac2{3}  \left( \ct\,\tilde{V}_{a1}  - \st\, \tilde{V}_{a2} 
\frac{\bar{\mt}(\mu_{SUSY})}{\sqrt{2}\sin\beta\,\mw} \right)^2
\frac{\mw^2}{\mst{1}^2} F_{7}^{(1)}(x_{{\tilde{t}_1}\,\chi^+_a}) \nonumber \\
&+&  \frac{K}{\cos\beta} \left(
\frac{\tilde{U}_{a2}\tilde{V}_{a1}\mw}{\sqrt{2} m_{\chi^+_a}}
\left[F^{(3)}_{7}(x_{\tilde{q}\, \chi^+_a})- 
\ct^2 \, F^{(3)}_{7} (x_{{\tilde{t}_1}\,\chi^+_a}) \right]
 \right. \non\\
&+&\left.  \left. \st \, \ct 
\frac{\tilde{U}_{a2}\,
\tilde{V}_{a2} \,\bar{\mt}(\mu_{SUSY})}{2\sin\beta \,m_{\chi^+_a}}
 F^{(3)}_{7} (x_{{\tilde{t}_1}\,\chi^+_a}) 
 \right) \right\}  \non \\
&+&  \sum_{a=1,2} \frac83 \left(\eta^{-\frac{14}{3 \beta_0}} -
\eta^{-\frac{16}{3 \beta_0}} \right) \left\{ 7 \rightarrow 8 \right\} \non \\
%%%%%%%%%%%%%%%%%%%%%%%%%%%%%%%%%%%%%%%%%%%%%%%%%
&+& \sum_{a=1,2} \left\{ - \frac2{3} \eta^{\frac4{\beta_0}}
 \left( \st\,\tilde{V}_{a1}  + \ct\, \tilde{V}_{a2} 
\frac{\bar{\mt}(\mu_{SUSY})}{\sqrt{2}\sin\beta\,\mw} \right)^2 
\frac{\mw^2}{\mst{2}^2} F_{7}^{(1)}(x_{{\tilde{t}_2}\,\chi^+_a})
\right. \nonumber \\
&-&  \frac{1}{\cos\beta}  \left[
\frac{\tilde{U}_{a2}\tilde{V}_{a1}\mw}{\sqrt{2} m_{\chi^+_a}}
\st^2 \left( K+\epsilon_b \tb \right)  \right. \non \\
&+& \left. \left.  \st \, \ct \frac{\tilde{U}_{a2}\,
\tilde{V}_{a2} \,\bar{\mt}(\mu_{SUSY})}{2\sin\beta \,m_{\chi^+_a}}
\left(K + \epsilon_b \tb \frac{\bar{\mt}(\mu_{W})}{\bar{\mt}(\mu_{SUSY})}  
\right) \right] F^{(3)}_{7} (x_{{\tilde{t}_2}\,\chi^+_a}) 
 \right\} \non \\
&+& \frac{\as (\mu_{W})}{4 \pi} 
\delta^S C_{7}^{(1)} (\mu_W) ~~.
\label{pippo}
\eea
In  eq.(\ref{pippo}) the term $\{ 7 \rightarrow 8 \}$ is the same as 
the one in the previous curly bracket, with the
$F_{7}$ replaced by  the $F_{8}$ functions. Also, $\beta_0= -41/6$ and
the term  $\delta^S C_{7}^{(1)} (\mu_W)$ is given in eq.(12) 
of ref.\cite{noi2}\footnote{In the published version of ref.\cite{noi2} 
there are typos corrected in the hep-ph archive version.}
with the following modifications to adjust for the resummation:
i) the  $\ln\mg^2/m_{\chi_j}^2$ 
in the functions $G_{7,8}^{\chi,1}$  in eqs.(15-17) of
\cite{noi2} must be replaced by  $\ln\mu^2_W/m_{\chi_j}^2$;
ii) the $\ln \mu^2_W/\mg^2$ in the functions $R_i$ in eq.(19) of 
\cite{noi2} must be dropped. 

The expression for the $C_8^\chi (\mu_W)$ 
coefficient can be obtained from eq.(\ref{pippo}) by dropping the 
$\{ 7 \rightarrow 8 \}$ term, by replacing the $\eta^{-\frac{16}{3 \beta_0}}$ 
factor that multiplies the
first curly bracket with $\eta^{-\frac{14}{3 \beta_0}}$, and by
substituting all the $F_{7}$ with the $F_{8}$ 
functions.

\section{Numerical analysis}
In this section we briefly 
illustrate the impact of the improved formulae, collected
in the previous section, in the calculation of
$B\to X_s \gamma$ branching ratio. 
We will consider here 
the case of a minimal supergravity model, in which gaugino and squark masses  
have common values $m_{1/2}$ and $m_0$, respectively, at the GUT scale. 
In the calculation of the branching ratio for $B\to X_s \gamma$ 
we include all  known perturbative and non--perturbative effects.
Next-to-leading order gluonic corrections to 
the Wilson coefficients at the electroweak
scale are included for the Standard Model \cite{nlo2,noi} and charged Higgs
contributions \cite{noi,BG}, 
together with the $\tan\beta$ enhanced terms introduced in
sect.~2. 
For what  concerns the chargino contributions, only the 
leading logs -- in the resummed form -- and the  $\tan\beta$ enhanced
terms discussed in sect.~3 
have been included beyond leading order. We estimate that
residual uncalculated next-to-leading order contributions, which  
do not present in this scenario any obvious enhancement, 
should be smaller than a few percent.
In the following we neglect $\tan\beta$ enhanced terms of electroweak
origin. 

\begin{figure}[h]%\vspace{-2.7cm}
\begin{center}
\mbox{\epsfxsize=13cm\epsffile{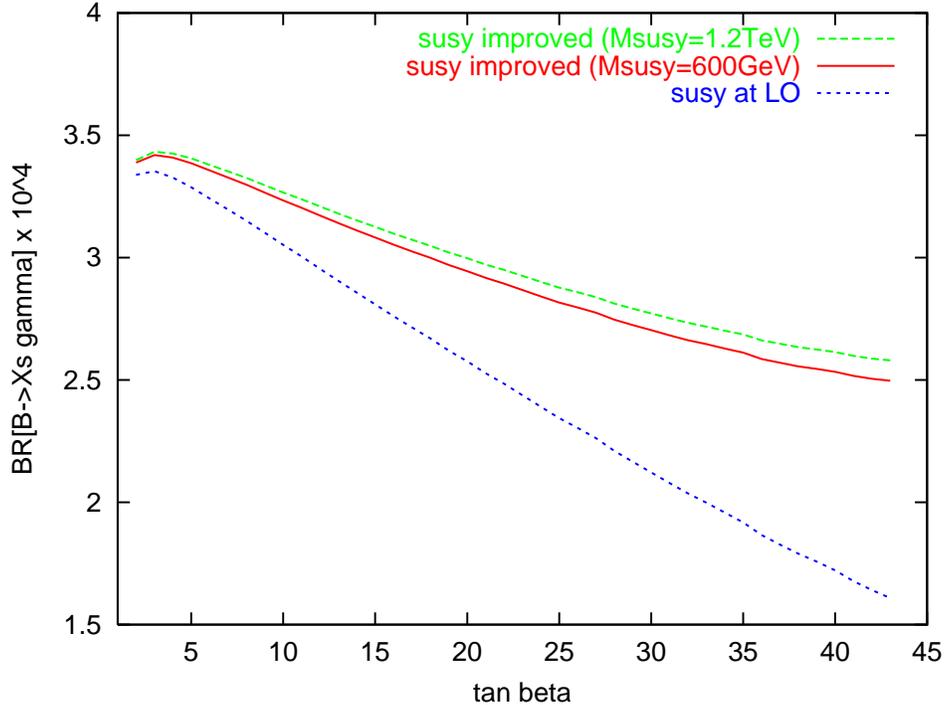}}
\end{center}
\vspace{-.5cm}
\caption{\sf Branching ratio for $B\to X_s \gamma$ in a minimal
  supergravity scenario with $m_0=600$ GeV, $m_{1/2}=$ 400GeV,
  $A_0=0$, and $\mu>0$ as a function of $\tan\beta$. 
The solid and dashed lines represent our
  improved framework for $\mu_{SUSY}=$ 600 GeV and 1.2 TeV, while the
  dotted line represents the results of the calculation with LO
  supersymmetric contributions.
\label{fig1}}
\end{figure}
In Figs.~2 and 3  we show the branching ratio for $B\to X_s \gamma$,
as a function of $\tan\beta$, with the parameter choice
$m_0=600$ GeV, $m_{1/2}=400$ GeV, and the common trilinear coupling at the
GUT scale $A_0=0$. The two figures correspond to the two possible signs of 
$\mu$,
with $|\mu |$ determined by the electroweak breaking condition.
This scenario is characterized by squark, charged higgs  and gluino masses
clustered
between 700 GeV and 1 TeV, with the charginos somewhat lighter.
Therefore we conservatively vary $\mu_{SUSY} $
between 
the two edges of the physical heavy-mass spectrum, and plot the
two extreme cases
$\mu_{SUSY}=600$ GeV (solid line) and 1.2 TeV
(dashed line). The dependence of the results on $\mu_{SUSY}$ appears to be
quite mild.
The improved predictions are then compared with the same calculation 
with supersymmetric contributions to the Wilson coefficients 
implemented at $\mu_W$ without improvements. 
The case of $\mu>0$ is characterized by destructive interference between 
Standard Model and chargino contributions. In this situation the improvements
have a significant effect. For $\tan\beta=40$ there is a 50\% enhancement
of the branching ratio. In the case of negative $\mu$, there is constructive
interference and the impact of the improved corrections is more limited. 

\begin{figure}[h]%\vspace{-2.7cm}
\begin{center}
\mbox{\epsfxsize=13cm\epsffile{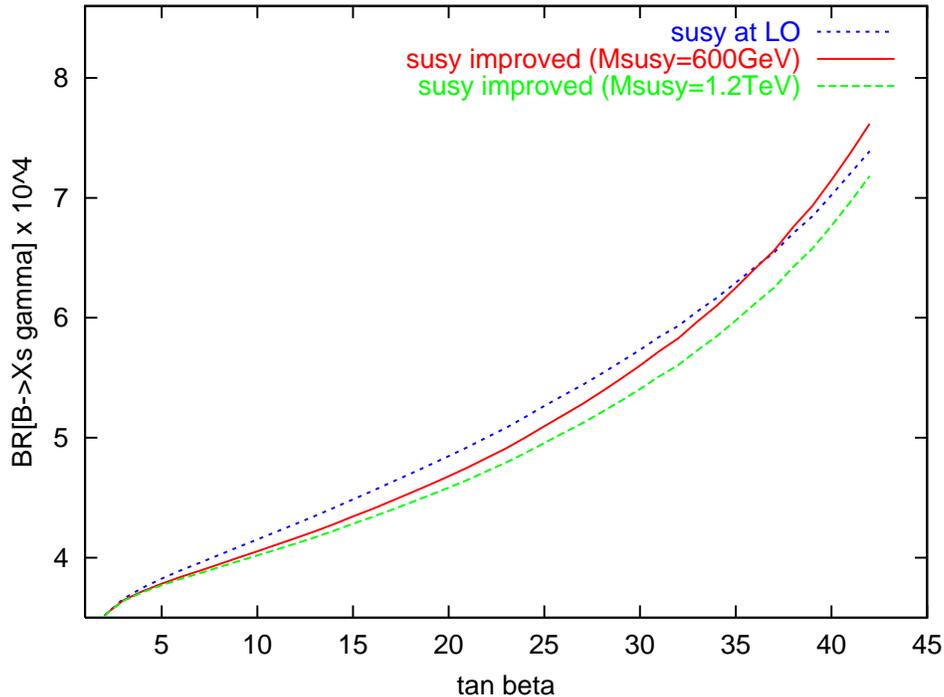}}
\end{center}
\vspace{-.5cm}
\caption{\sf Same as Fig.2 but with $\mu<0$.
\label{fig2}}
\end{figure}

We have  studied the scale ambiguity of the results and found
very small dependence on the matching scale $\mu_W$. The dependence on 
the $\mu_b$ scale is unchanged with respect to previous analysis \cite{noi}
and of the order of a few percent. Finally, the dependence on the
choice of $\mu_{SUSY}$ is shown in the plots and amounts  at most to
3\% (6\%) for positive (negative) $\mu$ in the branching fraction, 
for very large $\tan\beta$. 
An estimate of the residual theoretical uncertainty 
can be also obtained by noting that
in the context of the  next-to-leading calculation our improvements
can be implemented in two possible ways
which differ by higher order effects only.
One possibility is to use (29) in (24) and then identify the results as
improved leading order coefficients at $\muw$. The other option is to
identify the difference between the results of (24) and the leading order
 coefficients
as the $O(\alpha_s)$ or next-to-leading correction to the Wilson coefficient 
at $\muw$.  The difference between the two procedures is also $O(5\%)$.
 
We also want to
comment on an analysis which has appeared in ref.~\cite{deBoer}.
The authors use the next-to-leading calculation of ref.~\cite{noi2},
which assumes the hierarchy $\mst{1} \gg \mst{2}$, to compute the branching 
ratio
for $B\to X_s \gamma$ for the same scenario considered by us in Figs.\
2 and 3.
Unfortunately,
the mass hierarchy assumed in ref.~\cite{noi2} is not satisfied here, and
the approximate formulae miss the cancellation between the contributions from 
the two stop mass eigenstates. In this
situation the misuse of the result of ref.~\cite{noi2}
may lead to artificially large effects. Indeed, as shown in 
Figs.~2 and 3, our analysis does not confirm the large enhancement claimed
in ref.~\cite{deBoer}. 

\begin{figure}[h]%\vspace{-2.7cm}
\begin{center}
\mbox{\epsfxsize=13cm\epsffile{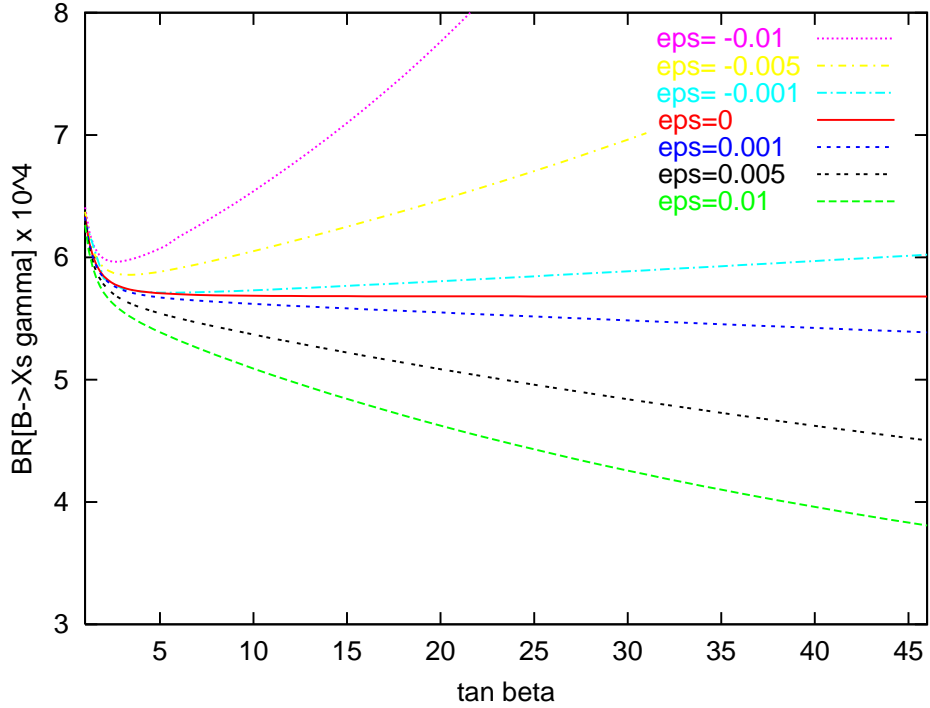}}
\end{center}
\vspace{-.5cm}
\caption{\sf Branching ratio for $B\to X_s \gamma$ in a two-Higgs
 doublet model with the charged-Higgs mass $M_H=150$~GeV, for
 different values of $\epsilon \equiv \epsilon_{b,t}=\epsilon_{b,t}^\prime$. 
\label{fig3}}
\end{figure}

Finally, let us consider the scenario in which the supersymmetric particles are
very heavy and let us focus on the contribution from the charged Higgs.
As explained in sect.~2, the existence of the
couplings $\epsilon_{b,t}$ and $\epsilon_{b,t}^\prime$ modifies the predictions
of what is generally called two-Higgs doublet model II even in the
decoupling limit. The impact of
the next-to-leading order calculation is illustrated in Fig.~4. This
figure shows the
branching ratio for $B\to X_s \gamma$ as a function of $\tan \beta$ for
$M_H=$~150~GeV. We have chosen $\epsilon=\epsilon_{b,t}=\epsilon_{b,t}^\prime$,
and varied its numerical value. 
For $\epsilon >0$ the next-to-leading
corrections reduce the leading order result. The effect at large $\tan\beta$ 
can be very significant and it  allows to consider 
values of the charged-Higgs mass, which 
were previously considered excluded, unless charginos and stops were
comparably light. The impact of the $\epsilon$ corrections on the
charged-Higgs mass lower limit from $B\to X_s \gamma$ is quantified in Fig.~5.
This figure is obtained by combining in quadrature theoretical
and experimental errors and using the most recent experimental
measurements for $B\to X_s \gamma$~\cite{cleo}.
Notice that for $\tan\beta =20$ values of the charged-Higgs
mass as low as 150~GeV  are allowed for 
%$\epsilon_{b,t}=\epsilon_{b,t}^\prime$ 
$\epsilon=10^{-2}$.

\begin{figure}[h]%\vspace{-2.7cm}
\begin{center}
\mbox{\epsfxsize=13cm\epsffile{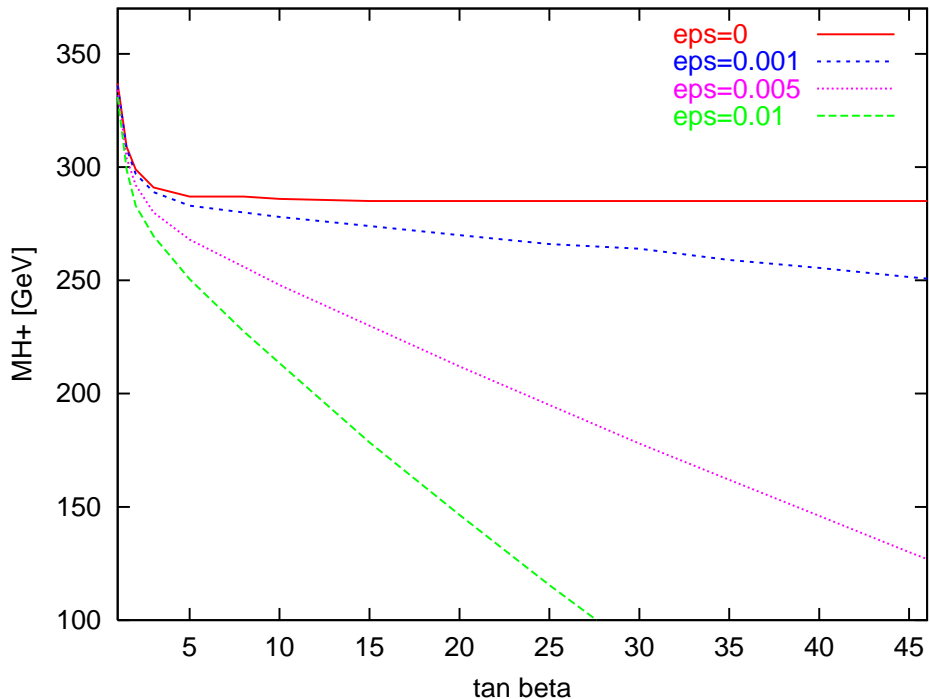}}
\end{center}
\vspace{-.5cm}
\caption{\sf Lower bounds on the charged Higgs boson mass 
obtained from the experimental measurement of  the 
branching ratio for $B\to X_s \gamma$ 
in a two-Higgs doublet model, for
 different values of $\epsilon\equiv\epsilon_{b,t}=\epsilon_{b,t}^\prime$ 
and as a function of $\tan\beta$. 
\label{fig4}}
\end{figure}

\bigskip

We are grateful to G.~Ganis and M.~Misiak 
for useful discussions and to R.~Rattazzi for very interesting 
and important remarks.

\end{document}